\documentclass[manuscript]{aastex61}

\usepackage{tikz, scalerel}

\definecolor{topcolor}{HTML}{00CC00}
\definecolor{secondcolor}{HTML}{FFFF33}
\definecolor{thirdcolor}{HTML}{FF8C1A}
\definecolor{fourthcolor}{HTML}{FF3333}
\definecolor{bottomcolor}{HTML}{999999}

\newcommand{\circledgreen}[1]		{\raisebox{-.35\height}{\begin{tikzpicture}\tikz[baseline=(char.base)]{\node[shape=circle,draw,fill=topcolor,inner sep=1pt] 		(char) {#1};}\end{tikzpicture}}}
\newcommand{\circledyellow}[1]	{\raisebox{-.35\height}{\begin{tikzpicture}\tikz[baseline=(char.base)]{\node[shape=circle,draw,fill=secondcolor,inner sep=1pt] 	(char) {#1};}\end{tikzpicture}}}
\newcommand{\circledorange}[1]	{\raisebox{-.35\height}{\begin{tikzpicture}\tikz[baseline=(char.base)]{\node[shape=circle,draw,fill=thirdcolor,inner sep=1pt] 		(char) {#1};}\end{tikzpicture}}}
\newcommand{\circledred}[1]	  	{\raisebox{-.35\height}{\begin{tikzpicture}\tikz[baseline=(char.base)]{\node[shape=circle,draw,fill=fourthcolor,inner sep=1pt] 	(char) {#1};}\end{tikzpicture}}}

\newcommand{\greenscore}[1]		{\begin{tikzpicture}\node[draw, fill=topcolor, circle, inner sep=1pt] (char) {#1};\end{tikzpicture}}
\newcommand{\yellowscore}[1]		{\begin{tikzpicture}\node[draw, fill=secondcolor, circle, inner sep=1pt] {#1};\end{tikzpicture}}
\newcommand{\orangescore}[1]	{\begin{tikzpicture}\node[draw, fill=thirdcolor, circle, inner sep=1pt] {#1};\end{tikzpicture}}
\newcommand{\redscore}[1]		{\begin{tikzpicture}\node[draw, fill=fourthcolor, circle, inner sep=1pt] {#1};\end{tikzpicture}}

\newcommand{\cmark}		{\raisebox{-.27\height}	{\circledgreen{	\scaleobj{0.70}{$\checkmark$}}}}
\newcommand{\xmark}		{\raisebox{-.20\height}	{\circledred{	\scaleobj{0.74}{-}}}}
\newcommand{\barriermark}	{\raisebox{-.21\height}	{\circledorange{	\scaleobj{0.76}{\vert}}}}
\newcommand{\peakmark}	{\raisebox{-.15\height}	{\circledyellow{	\scaleobj{0.85}{\sim}}}}
\newcommand{\mediummark}	{\raisebox{-.15\height}	{\circledorange{	\scaleobj{0.85}{\sim}}}}
\newcommand{\bottommark}	{\raisebox{-.15\height}	{\circledred{	\scaleobj{0.85}{\sim}}}}

\newcommand{\CR}{$\zeta_{\rm CR}$}
\submitjournal{ApJ}
\shorttitle{Atlas of CR-induced astrochemistry}
\shortauthors{Albertsson et al.}

\begin{document}
\makeatletter
\global\let\tikz@ensure@dollar@catcode=\relax
\makeatother
\title{Atlas of Cosmic ray-induced astrochemistry}
\correspondingauthor{Tobias Albertsson}
\email{albertsson@mpifr.de}
\author{Tobias Albertsson}
\affil{Max Planck Institut f\"ur Radioastronomie\\
Auf dem H\"ugel 69\\
53121 Bonn, Germany}
\author{Jens Kauffmann}
\affil{Max Planck Institut f\"ur Radioastronomie\\
Auf dem H\"ugel 69\\
53121 Bonn, Germany}
\author{Karl M. Menten}
\affil{Max Planck Institut f\"ur Radioastronomie\\
Auf dem H\"ugel 69\\
53121 Bonn, Germany}

\begin{abstract}
\noindent Cosmic rays are the primary initiators of interstellar chemistry, and getting a better understanding of the varying impact they have on the chemistry of interstellar clouds throughout the Milky Way will not only expand our understanding of interstellar medium chemistry in our own galaxy, but also aid in extra-galactic studies. 
This work uses the ALCHEMIC astrochemical modeling code to perform numerical simulations of chemistry for a range of ionization rates. We study the impact of variations in the cosmic-ray ionization rate on molecular abundances under idealized conditions given by constant temperatures and a fixed density of $10^4~\rm{}cm^{-3}$. As part of this study we examine whether observations of molecular abundances can be used to infer the cosmic-ray ionization rate in such a simplified case. We find that intense cosmic-ray ionization results in molecules, in particular the large and complex ones, being largely dissociated, and the medium becoming increasingly atomic. 
Individual species have limitations in their use as probes of the cosmic-ray ionization rate. At early time (<1 Myr) ions such as N$_2$H$^+$ and HOC$^+$ make the best probes, while at later times, neutral species such as HNCO and SO stand out, in particular due to their large abundance variations. 
It is, however, by combining species into pairs that we find the best probes. Molecular ions such as N$_2$H$^+$ combined with different neutral species can provide probe candidates that outmatch  individual species, in particular N$_2$H$^+$/C$_4$H, N$_2$H$^+$/C$_2$H, HOC$^+$/O, and HOC$^+$/HNCO. These still have limitations to their functional range, but are more functional as probes than as individual species. 
\end{abstract}


\section{Introduction} \label{sec:intro}
Astrochemistry is a complex web of processes that are hard to disentangle, as physical parameters affect the chemistry of the interstellar medium (ISM) in different ways. The cosmic-ray (CR) ionization rate, \CR, is the important ingredient that leads the way for ion-neutral reactions by ionizing neutral atoms and molecules. In particular, deep inside molecular clouds CR particles are the sole source of ionization, as the thick material shields the inner regions from ionizing UV photons \citep{2009ApJ...694..257I}. In extreme environments CR ionization can drive the formation of species but also destroy them as molecules are ionized and quickly dissociated due to a large amount of free electrons. 

CRs are predominantly charged particles, with protons being the dominating component, and smaller contributions from electrons, positrons, and bare nuclei of helium and other heavy elements \citep[see, e.g.][for a thorough review]{2012PhRvD..86a0001B}. 

But the origin of CRs is still not fully understood. It is impossible to directly pinpoint the point of origin as the charged CR particles\footnote{In the following we use the terms cosmic rays (CRs) and cosmic-ray particles synonymously.} are deflected and redirected by the Galactic magnetic field. Additionally, disentangling the direct CR contribution is complicated due to the interactions CR particles suffer on their way to Earth, and their interactions with X-rays. 

CRs can be studied with a variety of methods. Astrochemical methods, as discussed here, provide one opportunity. Specifically, these methods are sensitive to the ionization rate caused by CR interactions. Other methods are summarized by, e.g. \citet{2015ARA&A..53..199G}. It is, for example, possible to directly detect the protons, electrons, and other constituents of the CR sea in particle detectors. Unfortunately, the trajectories of CR particles with energies below a few GeV are heavily influenced by the solar wind. One must correct for this so-called solar modulation if one wishes to study these lower-energy CRs via direct detection. This possibly excludes CR data from \textit{Voyager 1}, which potentially has left the heliosphere. Alternatively, one can explore CRs via indirect methods, i.e., by studying the electromagnetic radiation CRs produce as they propagate through interstellar space. Synchrotron radiation from CRs moving in magnetic fields is one example. CRs also produce $\gamma$~rays when they interact with ISM particles: the collisions produce pions that decay in a variety of ways. The $\gamma$~ray observations allow us to study CRs down to energies $\sim{}1~\rm{}GeV$ \citep[e.g.][]{2014A&A...566A.142Y}.

CRs with energies $\lesssim{}0.1~\rm{}GeV$ are the primary agent responsible for the ionization in the dense sectors of the ISM that are shielded well from UV~radiation \citep[e.g., Fig. 14 of][]{2009A&A...501..619P}. Note that the properties of these CRs are poorly constrained by many of the aforementioned experiments: direct detection experiments and investigations of $\gamma$~rays can only provide reliable statements for energies $\ge{}1~\rm{}GeV$. This means that astrochemical studies that are sensitive to the ionization rate can deliver meaningful constraints on low-energy CRs, but it also means that CR-influenced astrochemistry is badly constrained by other observations, such as those of gamma-rays. 

The importance of these CR ionization events for initiating interstellar chemistry means that observations of molecules become useful probes for \CR. ISM ion-molecule chemistry begins with the formation of H$_3^+$, which is directly linked to the CR ionization rate as
\begin{eqnarray}
{\rm H}_2 + {\rm CR} 	&\Rightarrow& {\rm H}_2^+ + {\rm e}^{-} \\
{\rm H}_2^+ + {\rm H}_2 	&\Rightarrow& {\rm H}_3^+ + {\rm H} 
\end{eqnarray}
H$_3^+$ can then interact with neutral atoms and small molecules to form larger molecules, such as HCO$^+$, OH$^+$ and H$_2$O$^+$, which are thus indirectly linked to the CR ionization rate. For an overview of the ion chemistry in the interstellar medium, we refer the reader to \citet{2008ARAC....1..229S}. 

As such, observations of ions and other species are essentially indirect probes of the cosmic-ray ionization through its effects on the interstellar chemistry. By observing and interpreting the abundances of molecules in the ISM we can therefore constrain \CR. This can, for example, be done using statistical methods, such as $\chi^2$ minimization for fitting parameters. In the case of a few well-understood molecules with a simple chemistry, it is even possible to estimate the CR ionization rate directly from observed molecular abundances \citep[see, e.g.][]{1998ApJ...499..234C, 2010A&A...521L..10N, 2012ApJ...758...83I, 2015ApJ...800...40I}. 

But first we need to get a better understanding of CR-driven chemistry in the ISM by exploring how and which molecules are affected by the CR ionization rate the most. The aim of this paper is therefore to seek better probes that can improve future surveys of the CR ionization rate by studying the distinctive effect that it has on a selection of species often found in large surveys. To develop a first overview here we focus on a --- rather idealized --- situation in the parameter space of cloud structure. We consider molecules in a medium with constant gas and dust temperature at a density of $10^4~\rm{}cm^{-3}$ that develop out of some initial state assumed to be the same for all calculations. Reality is likely to be more intricate. However, our approach reduces the problem to a tangible one, and the calculations are certainly suited to revealing the astrochemical complexity encountered even under the most simple conditions.

In Section~\ref{sec:CRrate} we discuss previous observational measurements. In Section~\ref{sec:model} we present our model and a comparative model of derived measurement techniques. The results of the models are separated into discussions of the general trends of the increasing \CR~in Section~\ref{sec:chemsen}, and single species and species pairs as effective probes in Section~\ref{sec:single} and \ref{sec:pairs} respectively. We discuss our conclusions in Section~\ref{sec:conclusions}.

\section{Observational measurements}\label{sec:CRrate}
\begin{figure*}[!htb]
\centering
\includegraphics[width=0.99\textwidth]{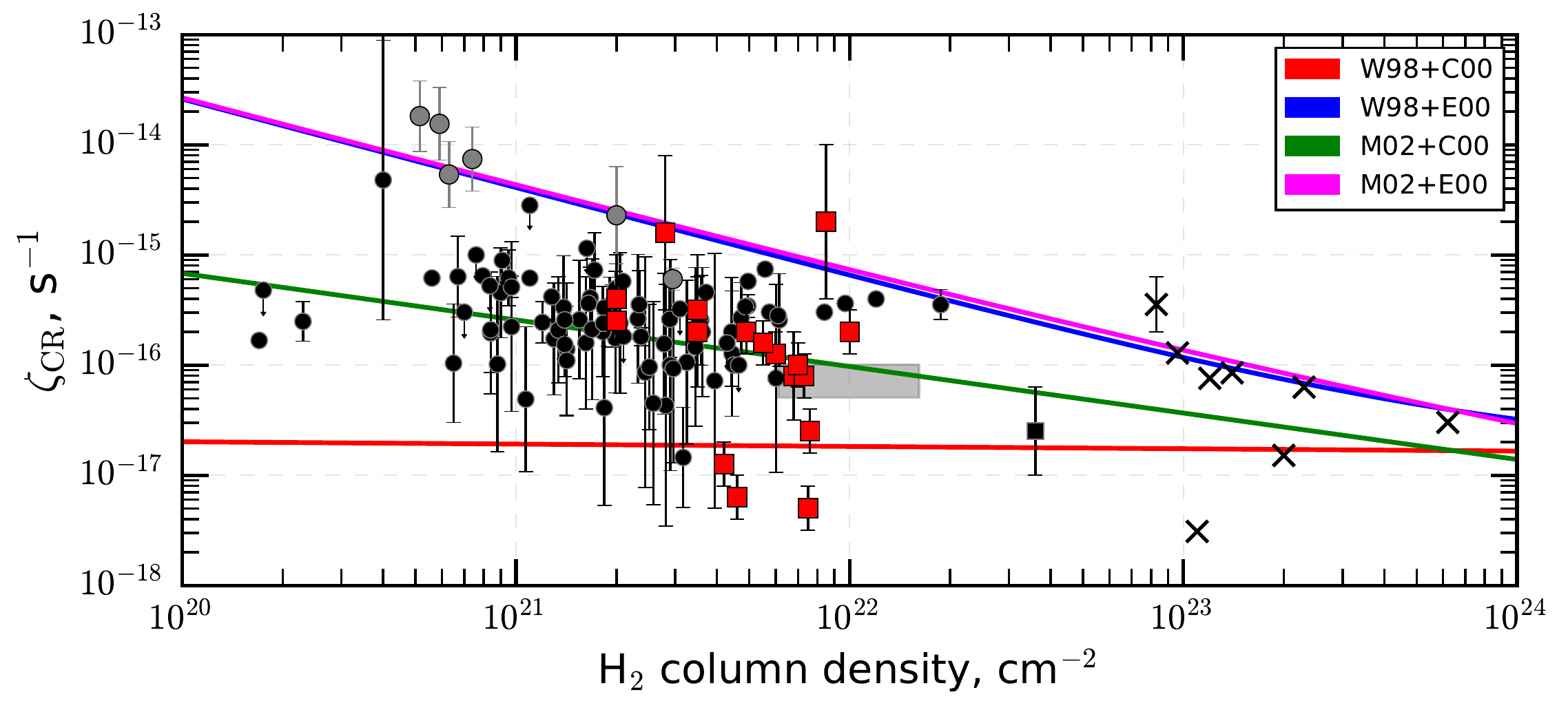}
\caption{Compilation of observed sources with measured H$_2$ column densities and $\zeta_{\rm CR}$. The symbols specify the type of object in the observations as follows: diffuse clouds ($\blacksquare$), low-mass cores ($\bullet$), and massive protostellar envelopes ($\times$). Data from diffuse clouds come from \citet{2002ApJ...577..221R}, and \citet{2007ApJ...671.1736I, 2015ApJ...800...40I}; low-mass cores are from \citet{1998ApJ...499..234C}, \citet{1998ApJ...503..689W}, \citet{2001Natur.409..159A}, and \citet{2007ApJ...664..956M}; massive protostellar envelopes are from \citet{1996A&A...310..315D}, \citet{2000ApJ...537..283V}, \citet{2000A&A...358L..79V}, \citet{2002A&A...389..446D}, and \citet{2008ApJ...675..405S}. The red squares are observations of low-mass cores by \citet{1998ApJ...499..234C}, the gray boxes are observations of diffuse clouds toward Sagittarius B2 by \citet{2015ApJ...800...40I}, and the gray box shows the range of values derived from the sample of low-mass cores by \citet{1998ApJ...503..689W}. The solid lines are from fits derived by \citet{2009A&A...501..619P} from proton \citep[W98]{1998ApJ...506..329W}, \citep[M02]{2002ApJ...565..280M} and electron \citep[C00, E00]{2000ApJ...537..763S} LIS data. This is an updated version of Figure~15 in \citet{2009A&A...501..619P}. 
} 
\label{fig:CRobs}
\end{figure*}

The rate of CR ionization events varies depending on the environment and its physical properties. In Figure~\ref{fig:CRobs} we plot CR ionization rates and H$_2$ column densities derived from observations toward diffuse clouds, low-mass cores, and protostellar envelopes. We also include the models derived by \citet{2009A&A...501..619P}, who considered two different determinations of the steady-state local interstellar spectrum (LIS) for CR protons by \citet[][``minimum"; W98]{1998ApJ...506..329W} and \citet[][``maximum"; M02]{2002ApJ...565..280M}, and two electron LISs by \citet{2000ApJ...537..763S}, C00 for their ``conventional" model C and E00 for their steeper (``SE") model. We combine the two proton and electron LIS results in the four lines in Figure~\ref{fig:CRobs}. 

Following \citet{1973ApJ...185..505H}, a CR ionization rate of $3\times{}10^{-17}~\rm{}s^{-1}$ is often adopted for dense cores in the solar neighborhood. One should, however, note that quite a range of values for the CR ionization rate has been reported in the literature \citep[e.g.][]{1998ApJ...499..234C, 1998ApJ...503..689W, 2007ApJ...664..956M}. In particular, exploration of the large sample studied by \citet{1998ApJ...499..234C} yields rates in the range $10^{-18}~{\rm{}to}~10^{-16}~\rm{}$ s$^{-1}$. Such estimates are subject to a range of uncertainties - of which some are examined in this paper. Still, Caselli et al. claim that their data are best explained by an actual variation in the CR ionization rate. If so, a variation of \CR~by two orders of magnitude in the solar neighborhood is certainly worthy of future investigations.

Many previous studies have sought to quantify the relationship between the CR ionization rate and measurables such as column densities of different ions. In Figure~\ref{fig:CRobs} we plot the results of observations for determining the CR ionization rate in different objects. Atomic and molecular diffuse clouds themselves lie in the range $10^{-16} - 10^{-15}$ s$^{-1}$, while dense molecular clouds and low-mass cores lie in the lower range 5$\times 10^{-18} - 1 \times 10^{-15}$ s$^{-1}$. 

\section{Model and measurement techniques}\label{sec:model}
One of the goals of this study is to present our adopted model and then compare it to equations derived for determining the CR ionization rate from observational quantities. 

\subsection{Model}
In this study we have utilized the gas-grain chemical model ``ALCHEMIC" developed by \citet{2010A&A...522A..42S}, who also presented a detailed description of the code and its performance. Below we give a brief explanation of the code. A UV photodesorption yield for surface species of 10$^{-3}$ is assumed \citep[see, e.g.][]{2009A&A...504..891O, 2009A&A...496..281O}. The self-shielding of H$_2$ from photodissociation is calculated by Equation 37 with \citet{1996ApJ...468..269D}, assuming a total extinction of 0.5 mag with a total column density $1.1 \times 10^{21}$ cm$^{-2}$ \citep{2009MNRAS.400.2050G}. The shielding of CO by dust grains, H$_2$, and its self-shielding, are calculated using the precomputed table from \citet{1996A&A...311..690L}. We use the chemical network from \citet{2013ApJS..207...27A}, which was updated for the ortho-para chemistry of H$_2$, H$_2^+$ and H$_3^+$ by \citet{2014ApJ...787...44A}. For initial abundances we adopt the ``low metal" initial abundances from \citet{1982ApJS...48..321G}, meaning that the medium begins as half molecular (due to H$_2$ being the dominant species), and completely neutral. We have also updated several tens of photoreaction rates from the original chemical model using the calculations of \citet{2006FaDi..133..231V}.


\begin{figure}[!hbt]
\centering
\includegraphics[width=0.46\textwidth]{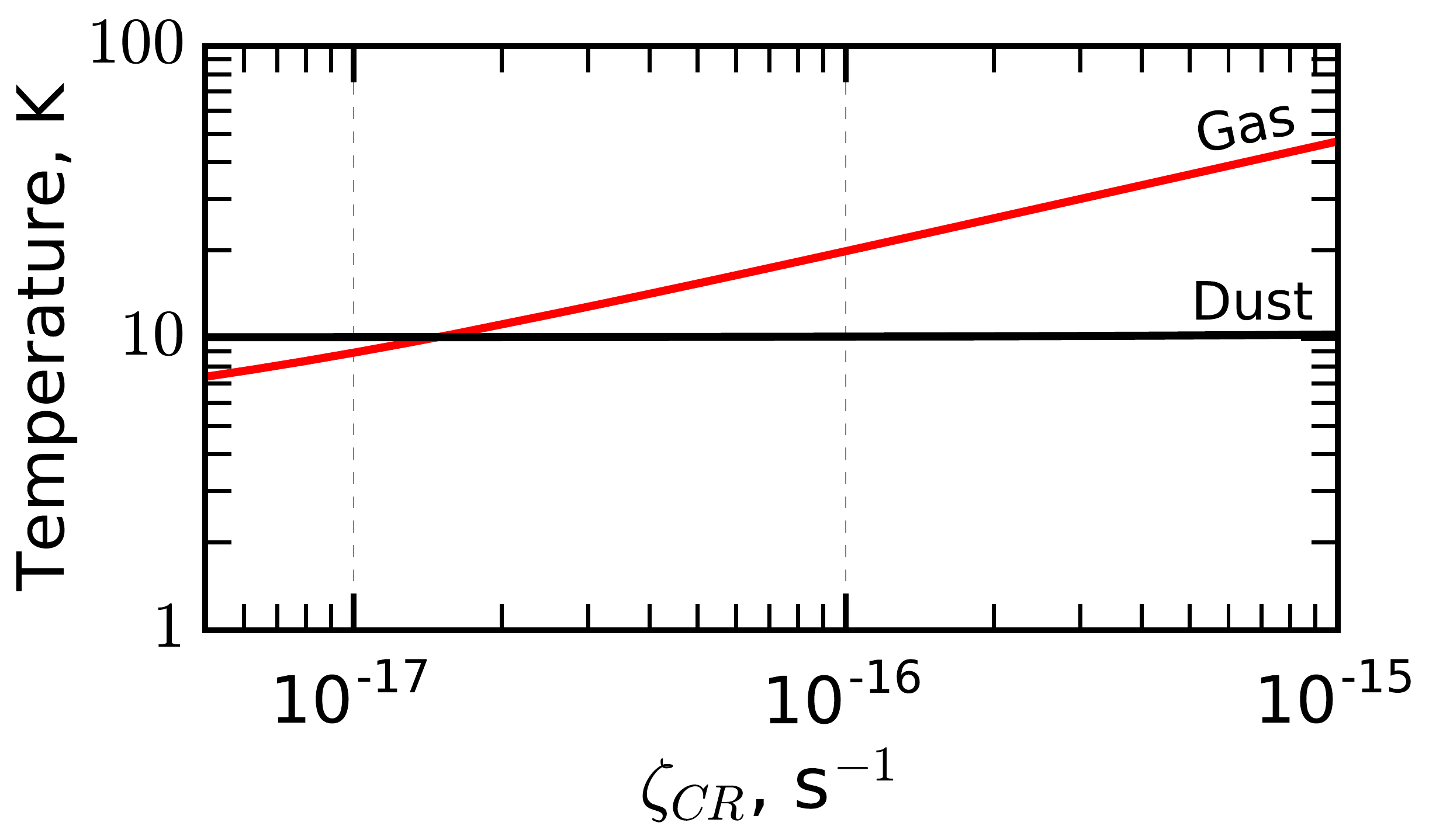}
\caption{Gas and dust temperature as a function of \CR, based on heating and cooling rates calculating using the procedure described in \citet{2001ApJ...557..736G}.
\label{fig:temperature}}
\end{figure}

\begin{deluxetable*}{ll}
\centering
\tablewidth{1.0\textwidth}
\tabletypesize{\small}
\tablecaption{Selection of observationally relevant molecules of primary concern in this paper. CH$_3$CCH is renamed to C$_3$H$_4$ in the adopted chemical network. \label{tab:selected_mols}}
\tablehead{
\colhead{Molecules}							&	\colhead{Reference}
}
\startdata
C, C$^+,$ CO, O												&	\citet{2001ApJ...557..736G} and \citet{2013ApJ...768L..34C}		\\
HNCO, HCN, HCO$^+$, HNC, N$_2$H$^+$, CH$_3$OH, CS, CN, CO		&	\citet{2014ApJ...788....4W}						\\
C$_3$H$_2$, CH$_3$CCH (C$_3$H$_4$), C$_4$H, NH$_2$D, SO			&	\citet{Frau2012}								\\
HC$_3$N, C$_2$H, SiO											&	MALT90 survey; \citet{2013PASA...30...57J}, 			\\
															&	\citet{2011ApJS..197...25F}, and \citet{2013PASA...30...38F}		\\
HOC$^+$														&	\citet{2015MNRAS.446.3842A}		\\
DCO$^+$														&	\citet{1998ApJ...499..234C}		\\
H$_3$O$^+$													&	\citet{2006AA...454L..99V}		\\
\enddata
\end{deluxetable*}

In Table~\ref{tab:selected_mols} we have compiled a list of observationally relevant molecules that we seek to investigate as possible probes. This selection is motivated by a number of studies that are broadly representative of observational investigations of molecules at millimeter wavelengths. Specifically, the line surveys by \citet{2011ApJS..197...25F, 2013PASA...30...38F}, \citet{Frau2012}, \citet{2013PASA...30...57J}, and \citet{2014ApJ...788....4W} give a broad overview of species that are easily detected in molecular clouds. We include additional molecules that are relevant for particular astrophysical reasons. These species are taken from the following studies: \citet{2001ApJ...557..736G} and \citet{2013ApJ...768L..34C}, who investigated cooling via spectral lines; \citet{1998ApJ...499..234C}, who introduced DCO$^+$ as a tool for CR determination; \citet{2015MNRAS.446.3842A}, who highlighted HOC$^+$ as an important ionized molecule; and \citet{2006AA...454L..99V}, who employed H$_3$O$^+$ as a probe of CRs in the Galactic Center region.

Based on Figure~\ref{fig:CRobs} we will investigate CR ionization rates in the range 5$\times 10^{-18}-$1$\times 10^{-15}$ s$^{-1}$ distributed in 25 logarithmically separated steps. Because CRs are heat sources, we include a routine to calculate the CR heating and cooling rates using the procedure described in \citet{2001ApJ...557..736G} to adjust the temperature based on the typical temperature of the modeled object. For these calculations we adopt a dust temperature of 10 K, consider negligible gas-dust coupling, and calculate gas temperature from balancing the heating and cooling rates based on the depletion factor at a density of 1$\times 10^{4}$ cm$^{-3}$. This gives us a improved estimate on the temperature structure of our object. Calculated gas and dust temperatures are shown in Figure~\ref{fig:temperature}. 

\subsection{CR estimation techniques}\label{sec:crest}
\begin{figure}[!htb]
\centering
\includegraphics[width=0.48\textwidth]{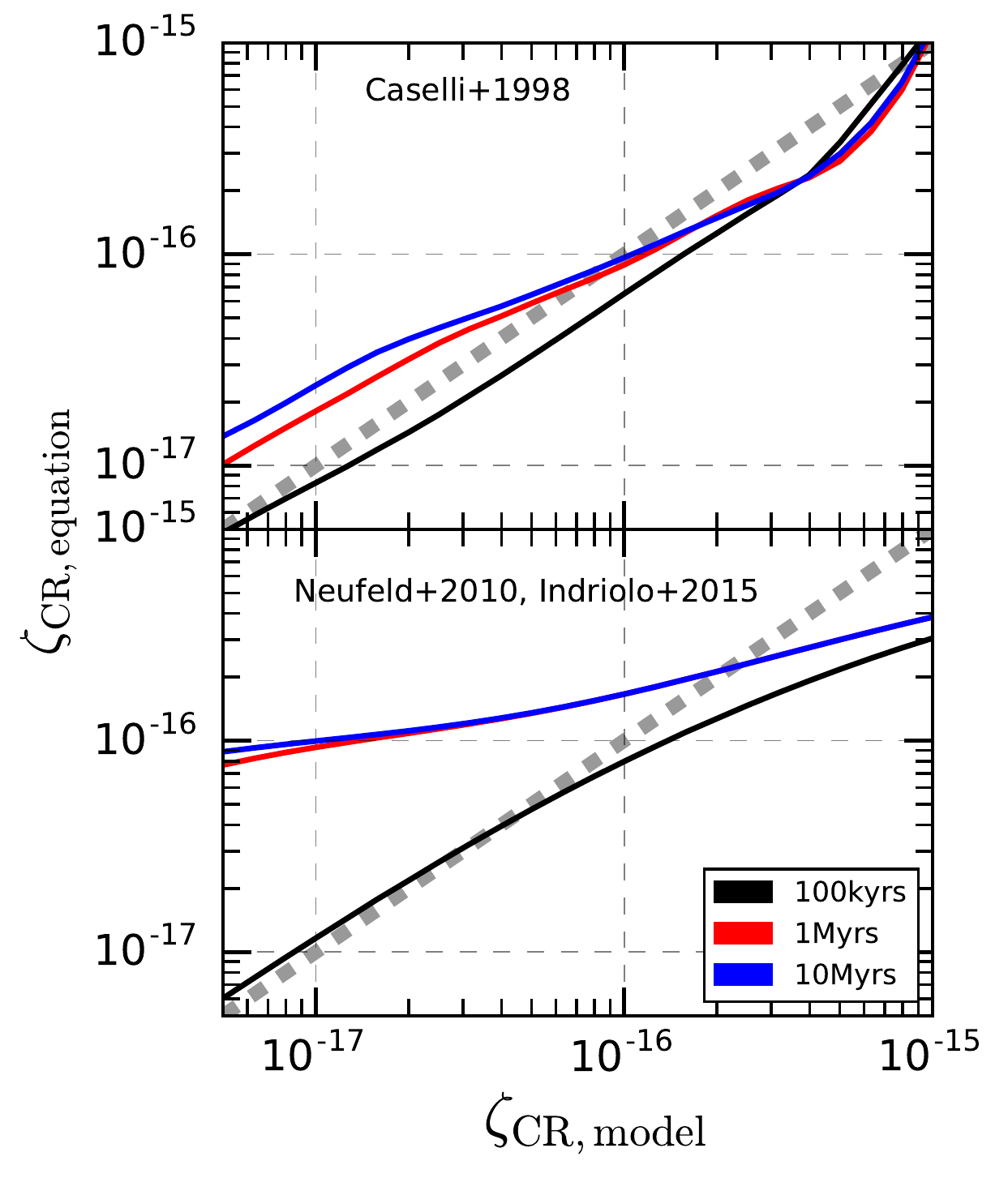}
\caption{Calculated CR ionization rates from relations derived by different studies vs. the actual modeled CR ionization rate. We adopt $\epsilon = 0.07\pm 0.04$ \citep{2012ApJ...758...83I}. The gray dashed line represents a perfect agreement between the CR ionization rate relations and the actual modeled CR ionization rate.}
\label{fig:CRcalc}
\end{figure}

So far, one of the most practiced methods for constraining \CR~is using ions and other species as indirect probes by interpreting their abundances based on our understanding of the interstellar chemistry. In Figure~\ref{fig:CRcalc} we plot the calculated versus the modeled CR ionization rate. Equations are based on studies by \citet{1998ApJ...499..234C} for dense clouds (top panel), and \citet{2010A&A...521L..10N} and \citet{2015ApJ...800...40I} for diffuse clouds (bottom panel). The dotted line represents a perfect agreement between the equations and our model. The relation by \citet{1998ApJ...499..234C} fits our model results quite well, within a factor 2--5. We also see changes between 1 and 10 Myr, but the chemistry does not change significantly after 10 Myr.

The equations from \citet{2010A&A...521L..10N} and \citet{2015ApJ...800...40I} have a much larger disagreement with the models for diffuse clouds. But if we exclude the lower end of the CR ionization rate, $\lesssim 10^{-16}$ s$^{-1}$ as CR ionization rates are higher than those in dense clouds, they agree within a factor of 2--5. It appears that there is not much variation already after 1 Myr for diffuse clouds. 

\citet{1998ApJ...499..234C} determined the ionization degree and cosmic ionization rate for 24 dark cloud cores based on observations by \citet{1995ApJ...448..207B}. Employing simple steady-state models for deriving HCO$^+$/CO ratios \citep[see, e.g.][]{1979ApJ...234..876W, 1982A&A...107..107G}, they derived the following relation:
\begin{equation}	\label{eq:caselli}
\zeta_{\rm CR}	=	\left[(k_1 + k_2) + \delta\right] n(e^-) \frac{n({\rm HCO}^+)}{n({\rm CO)}} \frac{k_3}{k_4}
\end{equation}
where $\delta$ is the reaction rate for H$_3^+$ (and H$_2$D$^+$) due to reactions with neutral species such as CO and O. \citet{1998ApJ...499..234C} estimated this as $\delta = 6.5\times 10^{-13} /$ f$_D$ cm$^3$ s$^{-1}$, where $f_D$ = [CO$_{\rm gas}$]/$\left([{\rm CO}_{\rm gas}]+[{\rm CO}_{\rm ice}]\right)$ is the depletion factor, i.e. the abundance fraction of gaseous CO over total (gas phase plus ice mantle) molecules. [X] denotes the absolute abundance of species X. The rates $k_i$ of other reactions included are as follows:
\begin{eqnarray}
&k_1&:	{\rm (o/p)\textnormal{-}H}_3^+ 	+	e^-		~~\Rightarrow	{\rm H}_2			+	{\rm H}		\label{eq:1}	\\
&k_2&:	{\rm (o/p)\textnormal{-}H}_3^+ 	+	e^-		~~\Rightarrow	{\rm H} + {\rm H} 	+	{\rm H}	\label{eq:2}	\\
&k_3&:	{\rm HCO}^+			  ~~~+	e^-		~~\Rightarrow	{\rm CO}			+	{\rm H}		\label{eq:3}	\\
&k_4&:	{\rm (o/p)\textnormal{-}H}_3^+	+	{\rm CO}	\Rightarrow	{\rm HCO/HOC}^+	+	{\rm (o/p)\textnormal{-}H}_2	\label{eq:4}
\end{eqnarray}
For the comparison we adopted a TMC 1-like molecular cloud model with $n_{\rm H}~=~10^4$ cm$^{-3}$, $A_{\rm V}$~=~10 magnitudes, $n(e^{-})$ calculated from the models, and compared our model to Equation~\ref{eq:caselli}. 

Modeling of dense clouds is easier, as we had high extinctions that block out UV photons and other additional processes, and we are left with CR photons as a dominant process. But some of the results presented here can also be transferred to diffuse clouds, and for comparison we also briefly discuss CR calculations in diffuse clouds. Compared to dark clouds, diffuse clouds have lower extinction values, which leads to higher effective CR ionization rates. Together with the high temperatures, this results in a different chemistry than that in dark clouds. More specifically, it leads to other ions dominating the chemistry. H$_3^+$ is not as abundant because of a high abundance of free electrons, which efficiently react with it and reduce its abundance. While HCO$^+$ is a strong probe for \CR~in dark clouds, OH$^+$ and H$_2$O$^+$ are two dominant ions that commonly are used as probes for \CR~in diffuse clouds. 

\citet{2010A&A...521L..10N} observed OH$^+$ and H$_2$O$^+$ from diffuse clouds along the sightline to the bright continuum source W49N. \citet{2012ApJ...758...83I} observed OH$^+$, H$_2$O$^+$, and H$_3^+$ in diffuse molecular clouds along sightlines near W51 IRS2. 

Following these results, \citet{2015ApJ...800...40I} examined observations of OH$^+$, H$_2$O$^+$ and H$_3$O$^+$ in diffuse clouds observed toward 20 sightlines with different Galactocentric radii with bright submillimeter continuum sources. These studies focused on the OH$^+$/H$_2$O$^+$ abundance ratio as a critical probe of the molecular fraction. Assuming that all OH$^+$ is destroyed by reacting with H$_2$ and by dissociative recombination at a rate equal to its formation rate (e.g. steady-state chemistry), they derived an equation for calculating \CR:
\begin{equation}	\label{eq:indriolo15}
\epsilon\zeta_{\rm CR}	=	\frac{n({\rm OH}^+)n(e^-)}{n_{\rm H}}  \left[\frac{k_3}{n({\rm OH}^+) / n({\rm H}_2{\rm O}^+) - k_4/k_1} + k_2 \right]
\end{equation}
where $k_j$ are the reaction rates for the following reactions:
\begin{eqnarray}
k_1:	{\rm OH}^+ 		+	{\rm H}_2			&\Rightarrow&	{\rm H}_2{\rm O}^+	+	{\rm H}	\label{eq:5}	\\
k_2:	{\rm OH}^+		+	e^-				&\Rightarrow&	{\rm products}					\label{eq:6}	\\
k_3:	{\rm H}_2{\rm O}^+	+	e^-				&\Rightarrow&	{\rm products}					\label{eq:7}	\\
k_4:	{\rm H}_2{\rm O}^+	+	{\rm H}_2			&\Rightarrow&	{\rm H}_3{\rm O}^+	+	{\rm H}	\label{eq:8}	
\end{eqnarray}
and $\epsilon$ is an efficiency factor introduced by \citet{2010A&A...521L..10N} that accounts for the fact that not every CR ionization event results in the net production of an OH$^+$ molecule \citep[e.g. $\epsilon = 0.07\pm0.04$; ][]{2012ApJ...758...83I}. For this comparison we assume a similar diffuse cloud model to \citet{2015ApJ...800...40I} with $n_{\rm H}$~=~35 cm$^{-3}$, $T=100$ K, $n(e^{-}) = 5.25 \times 10^{-3}$ cm$^{-3}$, $\epsilon = 0.07\pm 0.03$ and $A_{\rm V}$~=~1~magnitude. Because of the intense conditions, these environments are often predominantly atomic. Hence, we initiate our models with atomic abundances, including hydrogen. 

It is clear that the equation overestimates \CR. But this overestimation decreases with time and at higher modeled \CR~it even underestimates the rate. It shows the problem with assuming steady-state chemistry.  \citet{2012ApJ...754..105H} argued, for example, that steady-state is achieved on a timescale $\sim{}10^9~{\rm{}yr}/n(H)$, corresponding to $\sim{}30~\rm{}Myr$ at the density of $35~\rm{}cm^{-3}$ adopted here. It is conceivable that gas in the diffuse ISM is subject to relatively constant conditions over long periods of time. In that case estimates following the \citet{2015ApJ...800...40I} formalism do yield correct results. Still, the caveat of steady-state chemistry should be kept in mind. UV photons are contributing significantly to the production of electrons in diffuse clouds, and this is taken into account in our model, but not the equations.

While the two relations reproduce our model results quite well, there are still discrepancies. Our aim is to outline the use of molecular probes of $\zeta_{{\rm CR}}$ that are now routinely accessible with advanced facilities such as the Atacama Pathfinder Experiment (APEX) or the Atacama Large Millimeter/submillimeter Array (ALMA). In the next sections we will review and discuss the chemical species that are the most promising as probes of the CR ionization rate over the full ionization range and chemical ages in dense clouds. The identified list of species might also help with finding possible probes in circumstances where the standard probes are not available. 

\section{General trends}
\label{sec:chemsen}
\begin{figure*}[hbt]
\centering
\includegraphics[width=0.49\textwidth]{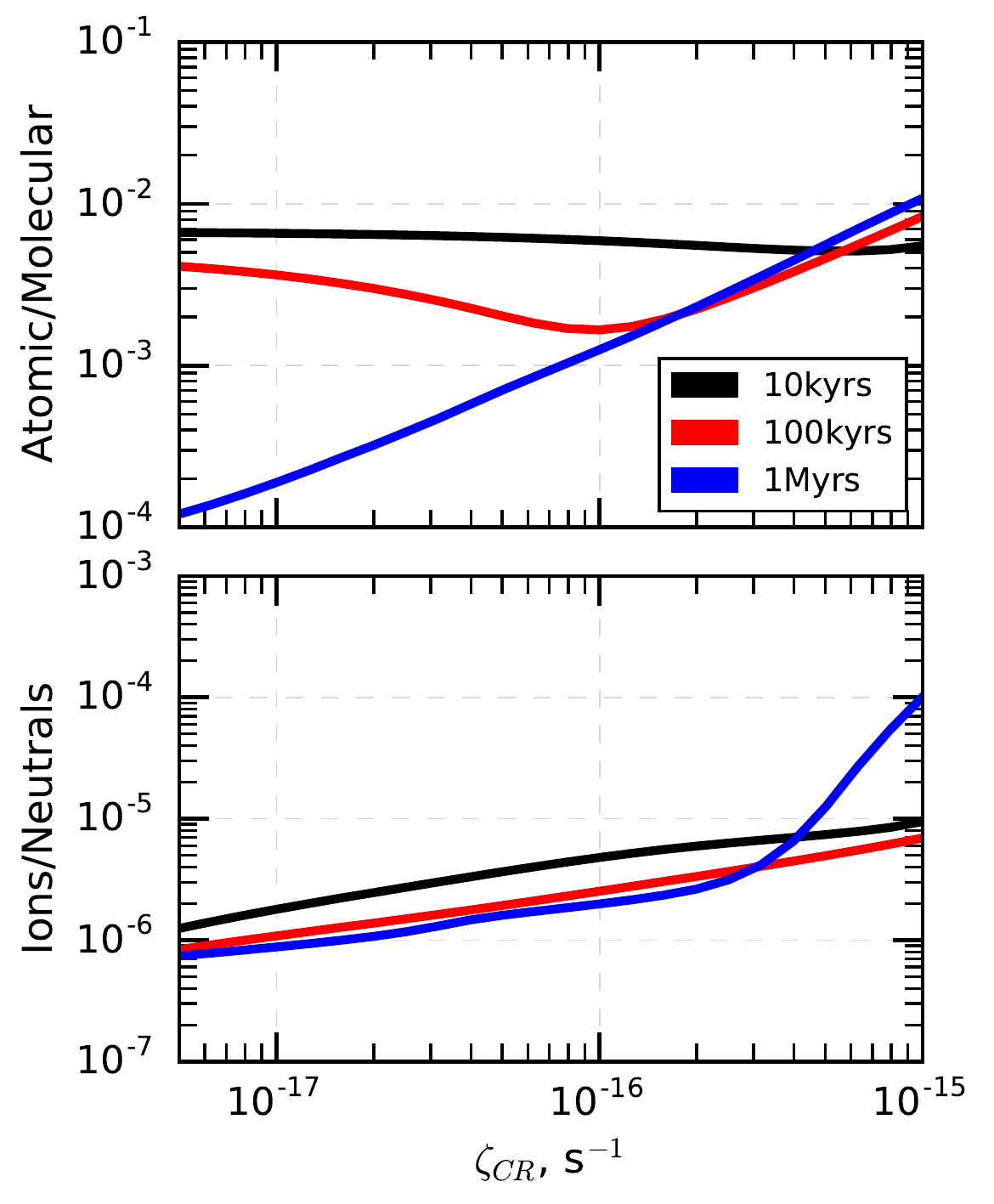}
\caption{The ratio between abundances for ions vs. neutral species is shown in the bottom panel, calculated by adding all species abundances that contain a charge (+/-). The figure shows the same general behavior as the electron abundance. The top panel shows the atomic vs. molecular (right) abundances as a function of \CR. 
}
\label{fig:ions_and_atoms}
\end{figure*}
We begin by looking at the overall trends in the chemistry to get an idea about the effects an increase in \CR~has on the chemistry of molecular clouds. Because our selection of species is small, we have calculated these trends based on the full chemical network, consisting of approximately 1300 species. In Figure~\ref{fig:ions_and_atoms} we look at the ratio of atomic versus molecular abundances (top panel) and the ionization degree of the material (bottom panel) as a function of \CR. 

The atomic versus molecular abundance ratio in the left panel is calculated as the ratio between the mass-weighted sums of the two groups: 
\begin{equation}
\sum_{i} m_{\rm atoms} \times n_{\rm atoms}/\sum_{j} m_{\rm molecules} \times n_{\rm molecules}\nonumber
\end{equation}
Up to 10 kyr the ratio is fairly flat and not affected by the increasing \CR. It is not until later that the effects starts to show. At 100 kyr the material has become overall more molecular, with the small exception for the highest \CR~$\sim 10^{-15}$ s$^{-1}$ for which molecular dissociation has already begun dominating the chemistry. At 1 Myr the trend from 100 kyr continues, and from here on the molecular dissociation stepwise begins dominating with time, which results in a monotonic increase in the ratio with \CR. As such, for old objects $\gtrsim$1 Myr, this could be a very powerful probe. It requires a detailed examination of the abundances of atoms versus molecules, which we leave for a future study. 

The effects of molecular dissociation are much clearer when we consider the abundance ratio between ions and neutral species, shown in the right panel of Figure~\ref{fig:ions_and_atoms}. Charged species can carry either a positive (+) or negative (--) charge, and we calculate the mass-weighted ratio as
\begin{equation}
\sum_{i} m_{\rm ions} \times n_{\rm ions}/\sum_{j} m_{\rm neutrals} \times n_{\rm neutrals}\nonumber
\end{equation}
The distribution clearly shows that ions are sensitive to an increase in \CR, as we see a shallow but clear trend of the ratio increasing with \CR. But it is the large increase at 1 Myr for \CR~$\gtrsim 3 \times 10^{-16}$ s$^{-1}$ that is the strongest sign of the molecular dissociation, where the number of atomic ions is greatly increased. Aside from that, we see that the ratios decrease overall with time, meaning that the material is, with time, slowly becoming more neutral: part of the atomic material is reforming neutral molecules. Though the trend increases for old objects at the highest \CR~$\sim 10^{-15}$ s$^{-1}$ and the trends are monotonic at every time step, the variance is very small. Again, this should be investigated by future studies.

\begin{figure*}[!bthp]
\centering
\includegraphics[width=0.99\textwidth]{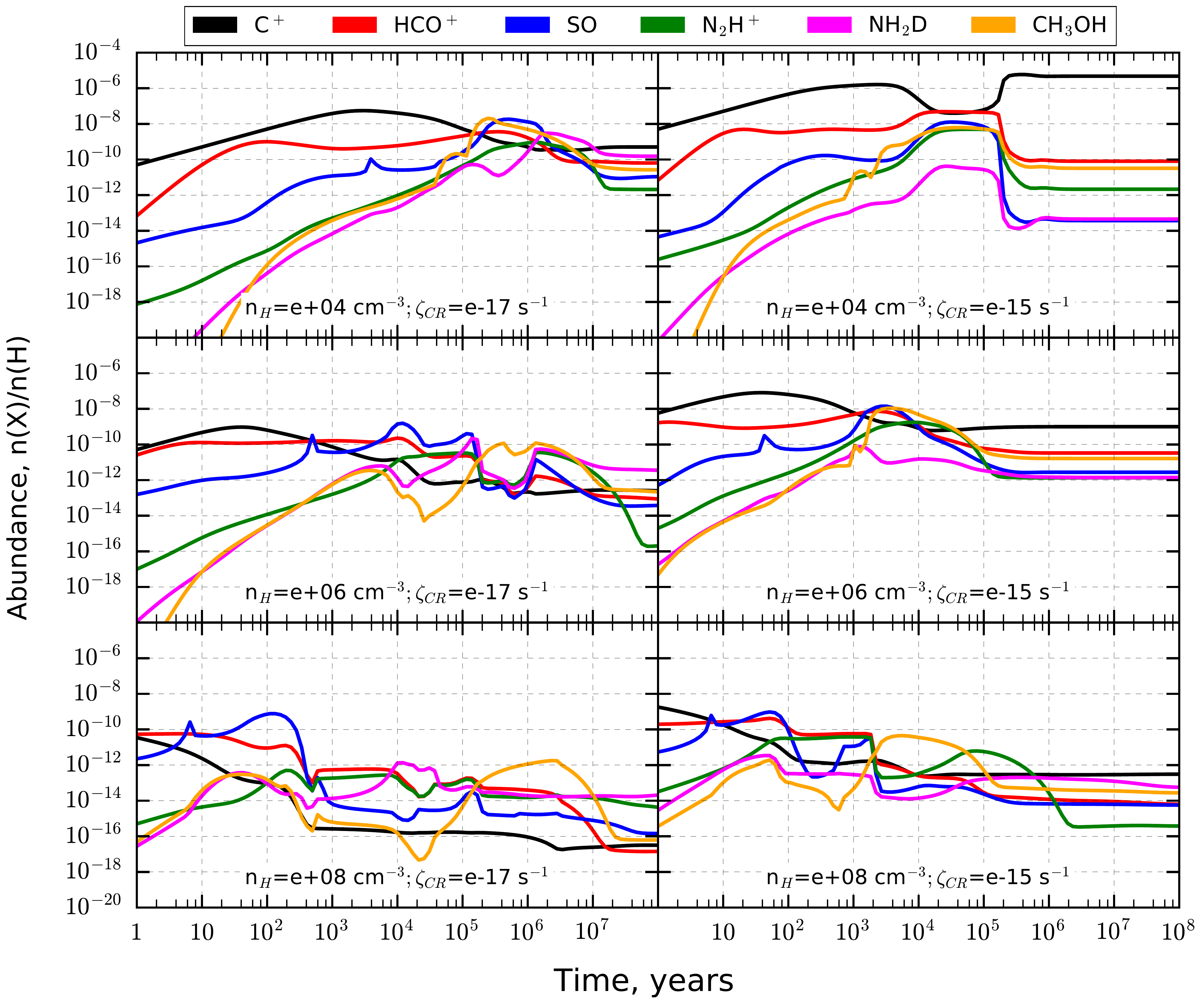}
\caption{Abundances for a selection of the most commonly observed molecules as a function of time, plotted for different densities and \CR. 
}
\label{fig:timedens}
\end{figure*}

We will also look at the effects of an increase in both density and time. Because temperature is directly linked to \CR~through the calculated cooling and heating rates, we do not investigate this parameter separately. 

In Figure~\ref{fig:timedens} we plot abundances for a variety of commonly observed molecules as a function of time for three different densities (rows; $10^4$, $10^6$ and 10$^8$ cm$^{-3}$) and two different CR ionization rates (columns; $10^{-17}$ s$^{-1}$ and $10^{-15}$ s$^{-1}$). 

Increasing the density has the strongest effect on the chemical time scale. In Figure~\ref{fig:timedens} we see the abundances increasing, in some cases monotonically, until they reach a peak value. This peak moves to earlier times with increasing density. After this point, the chemistry becomes more complex. Compared to previously only being dominated by ion--neutral reactions, surface chemistry and neutral--neutral reactions begin to play a significant role. An increase in small ions, in particular atomic ions, also affects the chemical equilibrium values. 

With time, often after approximately 1 or 10 Myr, depending on the ionization rate, the system reaches an equilibrium state. The equilibrium state of the species' abundances largely depends on the adsorption of species onto grain surfaces, which becomes more efficient with higher densities and \CR. The effect from an increase in \CR~is indirect, as it also increases the temperature, and thus the amount of material desorbed into the gas phase for some species, but also increases the amount of ions, which has significant effects on the chemistry. 

In comparison, \CR~can have very significant influence on the chemistry, in particular on the steady-state abundances. But density has a similar effect between low and high \CR. That means that density only competes with time to effect the chemistry, and its effect is much less significant. We conclude that density is not as important as time for the chemical evolution, and in order to simplify the analysis, we exclude it from the study by keeping it constant at $n_{\rm H} = 10^{4}$ cm$^{-3}$. 

\section{Individual tracers}\label{sec:single}
Here, we look at the overall trends in abundances for the species listed in Table~\ref{tab:selected_mols}. Species must fulfill a number of conditions to be useful as probes of the CR ionization rate. First, molecules must be sufficiently abundant to be observable; in our case we consider a relative abundance of $10^{-15}$ to be the limit of observability. Second, ideal probes must show a monotonous increase or decrease of their abundances versus \CR. Third, these abundance variations as a function of \CR~are ideally large so that differences can easily be observed. While the species that we consider in Table~\ref{tab:selected_mols} have been proven to be observable, and hence fulfill the first criterion, we will take into account the two other criteria. 

\subsection{Probe score}\label{sec:score}
The usefulness of an individual or pair of species will be discretized to a probe score, ranging between 0 and 1, based on the above criteria. The score is defined by the multiplication of two factors; the fractional range in \CR~where the species can be used as a probe, and the logarithm of the maximum abundance variations. By multiplying the two factors, which we have named as the $functional$ and the $variation$ factors, respectively, we calculate the probe score. 

The functional factor will be calculated based on the most dominant trend between positive and negative trends. In order to do this, we need to determine trends in the data and as such, need to define what constitutes a significant variation. All CR ionization events will not result in the production of the target species, represented by $\epsilon$ for the case of OH$^+$ in Section~\ref{sec:crest}. We consider a conservative effectiveness of 50\%. This means that for an increase in \CR~by one order of magnitude, abundance should increase by at least a factor of 5, and anything less signifies a $plateau$. After plateaus have been identified we use Theil-Sen regression to determine the remaining trends in the distribution. This is a robust regression method from the python package \textsc{scikit-learn} that ignores outliers. 

\pdfimageresolution=10
\begin{figure*}[!tb]
\centering
\includegraphics[width=1.00\textwidth]{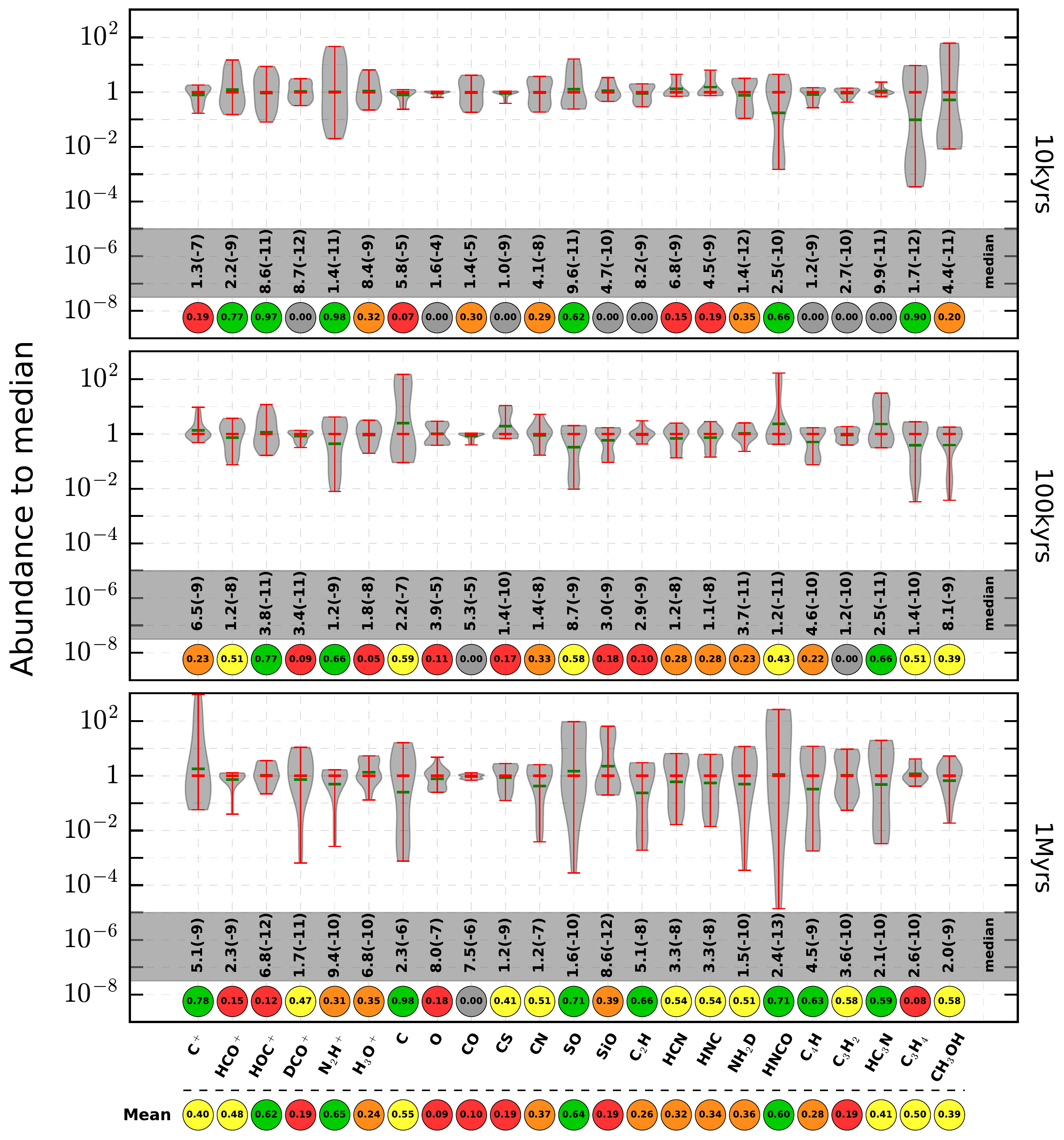}
\caption{Violin plots showing ranges of abundances for essential molecules after 0.01, 0.1, and 1 Myr. Values are normalized to each molecule's median abundance, which better represents the abundance variation between species, and for comparison the median abundance is shown near the bottom. Mean values are shown with a green line and the probe scores are calculated from the different quartiles of the total distribution of probe scores. The green scores represent the top quartiles with scores $\geq 0.59$, the yellow scores represent the second quartiles with scores $> 0.39$, the orange scores represent the third quartile with scores $> 0.20$, the red scores represent any non-zero values, and the gray scores represent a fully plateaued distribution.
}
\label{fig:violinplots}
\end{figure*}

A plateau limits the functional range of a probe, but a repeated switch between positive and negative trends will also have a great impact on how useful a probe is. If it switches once, we get a bow-shaped distribution, and as long as we can clearly distinguish between the positive and negative sides of the distribution, the species is still useful as a probe. However, if the trend changes more than once, it severely reduces how useful the species is as a probe, and in these cases we therefore halve the probe score.

The variation factor in the probe score is defined as the logarithm of the ratio between the maximum and the minimum abundance, $R =$ log$_{10}$~(max[X]/min[X]), for species X. Abundance variations will be restricted, as variations beyond a certain factor will not have a significant impact. In this study we consider this limit to be a factor of 100, based on our observational experience. We still want to be able to separate and rank species that have extremely large abundance variation. For this reason the variation factor will be linear with \CR~up to a factor 0.1 below the maximum allowed variance. After that, we adopt a relation $\propto x/(1+x)$, where $x$ is the logarithm of the abundance variation. Finally, the $variation$ factor is normalized. 

\subsection{Survey of observable species}
We start with an overview of our selection of species. In Figure~\ref{fig:violinplots} the abundances are shown separately for 0.01, 0.1, and 1 Myr (top to bottom) in so-called ``violin plots". Abundances are shown relative to their median values and for comparison the mean abundance is indicated with a green line. The median values of each species are printed at the bottom of each panel. The gray shapes show the probability density function, a non-parametric estimate of the probability distribution, determining the probability of specific abundances to be found at certain abundance ranges. 

The probe scores of individual species are shown at the bottom of each panel. They are color-coded based on the four quartiles of the distribution of calculated probe scores for all species in Table~\ref{tab:selected_mols} and the three investigated time steps. We split calculated probe scores into quartiles, each containing a fourth of the total probe scores. The green probe scores represent the top quartile with values $\geq$0.59, the yellow scores represent the second quartile with scores $\geq$0.39, the orange scores represent the third quartile with scores $\geq$0.20 and the red scores represent with scores >0.0. Cases for which there is no significant variation in the abundances over the complete \CR~range are flagged as plateau distributions and are represented in gray with a probing score of 0. 

The species in Figure~\ref{fig:violinplots} are arranged such that ions are shown first, followed by neutral molecules, both with an increasing number of atoms. Though ions are strongly connected to \CR, their probe scores are telling a different story, with many examples of having low probe scores. No ions show a persistently high score. This is due to a mix of small abundance variations and sensitivity to the increasing electron density that follows the increase in \CR. N$_2$H$^+$ and HOC$^+$ both stand out as great probes at times <1 Myr, and to some lesser extent also HCO$^+$, but at 1 Myr C$^+$ is the only useful probe. 

Amongst the neutrals, the trends vary a lot, and we find good probe scores for all different types and sizes of species. Amongst the atoms, C stands out as an excellent probe at 1 Myr, just like C$^+$, but at earlier times its probe scores also suffer from small abundance variations. Amongst diatomic species, SO is the sole species that shows promise, and in fact at all three time steps. The triatomic species all have below average scores, but amongst the more complex species we find several probes, and amongst these, HNCO and C$_3$H$_4$ are the most promising. 

\subsection{Selected tracers}\label{sec:ind_tracers}
\begin{figure*}[!htb]
\centering
\includegraphics[width=1.00\textwidth]{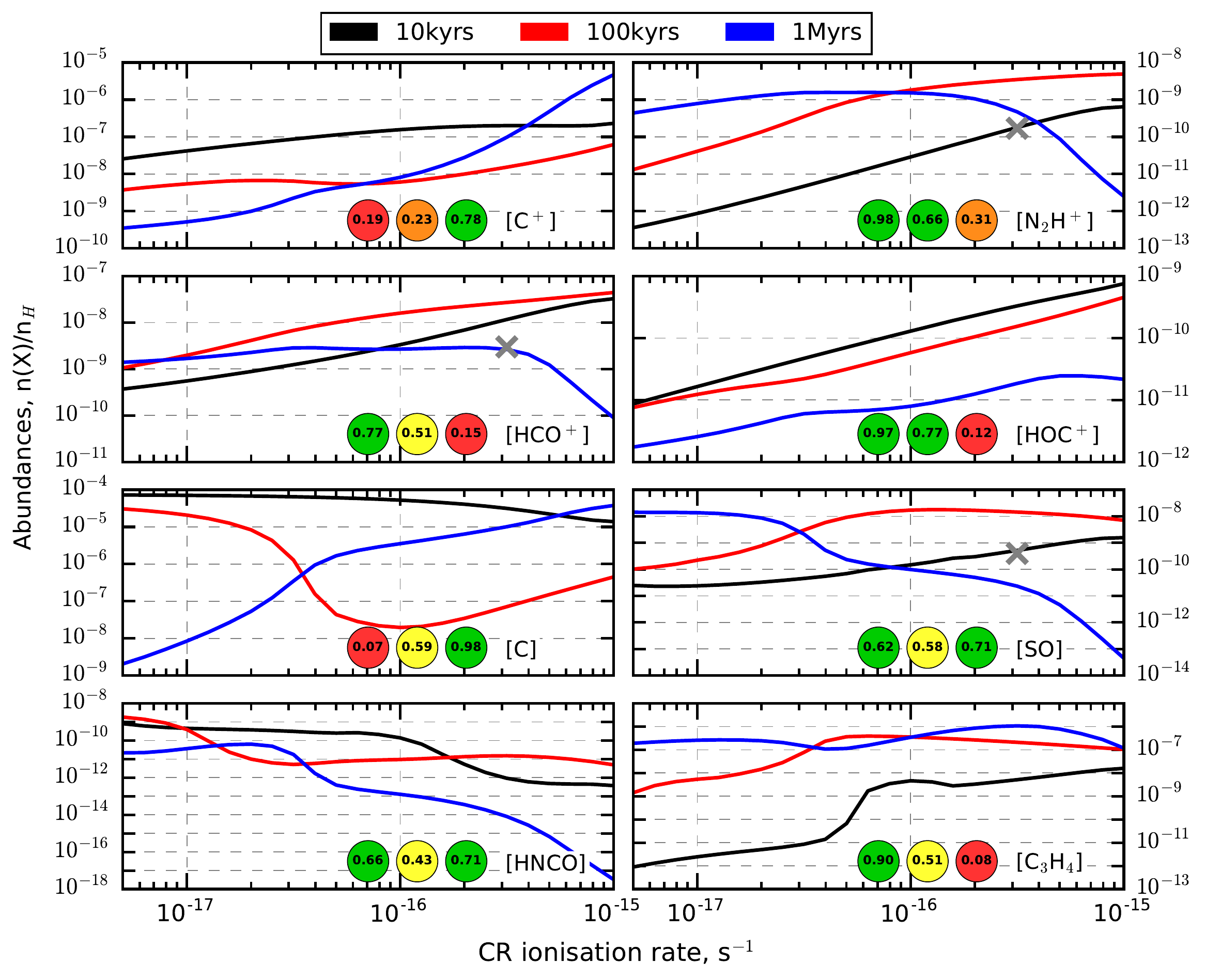}
\caption{Abundance as a function of $\zeta_{CR}$ for differently behaved molecules at times 0.01 (black line), 0.1 (red line), and 1.0 Myr (blue line). The green scores represent the top quartiles with scores $\geq 0.59$, the yellow scores represent the second quartiles with scores $> 0.39$, orange for third quartile with scores $> 0.20$, and the red scores represent any non-zero values.
}
\label{fig:showcase}
\end{figure*} 
As such we have identified eight promising probe candidates that we will take a closer look at by dissecting their chemistry and evaluating their fit as probes of \CR. In Figure~\ref{fig:showcase} we illustrate variations in abundances for these selected species and in the following section individually discuss their chemistry. In Table~\ref{tab:ind_regions} we summarize the regions where the individual species are useful as probes for \CR. 

We compare our model results with the observations of the starless core L1498 by \citet{2006A&A...455..577T}, in order to ascertain the accuracy of our numerical models. Tafalla et al. simultaneously observed N$_2$H$^+$, HCO$^+$ and SO from our list of species. We have marked the observed abundances in Figure~\ref{fig:showcase} with $\times$, adopting \CR~=~$3\times10^{-16}$ s$^{-1}$ derived by \citet{1998ApJ...499..234C}. The comparison points to a chemical age of $\sim$10 kyr or 1 Myr. These very different ages illustrate the uncertainties involved when only using a small number of species. The latter is closer to the derived lower age limit by \citet{2013A&A...559A..53M} of a few hundred-thousand years. Discrepancies are likely due to Maret et al. adopting the density structure from \citet{2004A&A...416..191T} instead of a single density point as in this paper. 

Atomic ionized carbon (C$^+$) does not increase strongly enough at 10 kyr, and at high \CR~it ends with a plateau. At 100 kyr molecular dissociation starts to affect the abundances, resulting in a wavy pattern during this transitional period. It is not until $\sim 1$ Myr that this ion becomes a promising probe. But once again, due to molecular dissociation, its abundance variation is approximately five orders of magnitude. 

N$_2$H$^+$ is linked to \CR~through its formation from H$_3^+$ reacting with N$_2$. At late chemical ages ($\approx 1$ Myr) it slowly reaches a peak abundance at \CR $\sim 10^{-16}$ s$^{-1}$, after which the abundance quickly drops several orders of magnitude due to molecular dissociation. This limits its usefulness as a probe at late times, which we see reflected in its probing score, while at earlier times it is a great probe with monotonic and large increasing trends. 

HCO$^+$ is a commonly used probe for \CR. But as we see in the second row of Figure~\ref{fig:showcase}, its isomer HOC$^+$ proves to be a much more consistent probe. At earlier times $< 1$ Myr, both HCO$^+$ and HOC$^+$ show a strong increase with \CR. It is at 1 Myr that the two ions differ. While HCO$^+$ suffers greatly from molecular dissociation, and at higher \CR~there is also an increase in o-H$_2$, which converts HCO$^+$ to HOC$^+$. As such, HCO$^+$ has a plateau until \CR~$\sim 3\times 10^{-16}$ s$^{-1}$ when the dissociation quickly drops abundances. Meanwhile HOC$^+$ benefits from a small but monotonic increase in abundance and only suffers from a plateau at \CR~$\sim 10^{-15}$ s$^{-1}$. As such, HOC$^+$ is clearly a more capable \CR~probe. Note that the abundance of HOC$^+$ is lower than HCO$^+$ by a factor of 10--100 depending on the time and \CR. This, of course, makes observations of HOC$^+$ more challenging than observations of HCO$^+$. 

Atomic carbon (C) shows a wide range of behaviors over the three time steps. At 10 kyr it shows a slow decrease in abundance, so slow that it is considered a plateau up until \CR~$\approx 10^{-16}$ s$^{-1}$. The decrease in the C abundance is much more significant at 100 kyr, dropping approximately three orders of magnitude until \CR $\gtrsim 10^{-16}$ s$^{-1}$, after which molecular dissociation slowly begins to dominate and increases the C abundance. This results in a much higher probe score, but because of the arc-shape, it is a functional probe over a very limited range in \CR. At 1 Myr the molecular dissociation dominates the chemistry of atomic carbon, resulting in a monotonic positive trend. Compared to C$^+$, the two are great probes at 1 Myr, but atomic carbon is also a useful probe at 100 kyr.

SO has a clear positive trend at 10 kyr, but at 100 kyr it suffers from a plateau at \CR$~\gtrsim 10^{-16}$ s$^{-1}$, again due to molecular dissociation. At 1 Myr this trend also spreads to lower \CR, but the abundance is monotonic aside from a brief plateau at low \CR~$\lesssim 10^{-17}$ s$^{-1}$. As such, SO is showing promise at all time steps, aside from a restricted functional range at 100 kyr. 

HNCO is a very peculiar case. At 10 kyr it is slightly restricted by a small plateau, but at 100 kyr it is restricted to only being useful up to \CR~$\sim 3\times 10^{-17}$ s$^{-1}$. The peculiarity arises at 1 Myr, when its abundance drops by almost eight orders of magnitude. This feature alone makes it interesting as a probe, in particular for more distant or diffuse targets. Reactions with H$^+$ and He$^+$ split the molecule into smaller parts, which are not so easy to reform. Considering that uncertainties in observational measurements increase for distant targets, HNCO shows great promise due to its large variations for for chemical ages $\gtrsim 1$ Myr. The large variations means it is not as sensitive to these errors as other probes and we can get more precise estimates of \CR. 

The final example is C$_3$H$_4$, another molecule primarily formed on the surface from C$_3$ being hydrogenated. At 10 kyr and 100 kyr, we see a sharp increase in the abundance at \CR $\sim 5\times 10^{-17}$ s$^{-1}$ that is most prominent at 10 kyr. This is caused by the sudden increased inflows of carbon and hydrogen atoms on the dust surfaces and by the dissociation of H$_2$ and longer carbon chains C$_{\rm m}$ for m$> 2$. Despite this bump, C$_3$H$_4$ shows an almost monotonic increase at 10 kyr, while at 100 kyr it is much more restricted at \CR~$\gtrsim 10^{-16}$ s$^{-1}$ due to reactions with molecular ions such as H$_3$O$^+$, H$_3^+$ and HCO$^+$. Finally, at 1 Myr the molecule shows fluctuations in its abundance, similar to a sinus curve, which is due to destruction from reacting with molecular ions at low \CR, while at \CR~$\gtrsim 5\times 10^{-17}$ s$^{-1}$ it begins to reform into C$_4$H$_2$ instead thanks to the strong increase in atomic carbon. Compared to HNCO, C$_3$H$_4$ reforms much more efficiently. This is because it reacts with ions produce molecular ions such as C$_3$H$_{\rm m}^+$, with $m \geq 2$, which can quickly reform into C$_3$H$_4$. However, in this case it is a drawback, keeping its abundance steady and the molecule not useful as a probe for older objects $\gtrsim 1$ Myr. 

\begin{table}[!tb]\normalsize
\centering
  \begin{tabular}{cccc}
	\hline	\hline\\[-1.4cm]
Species			&	\multicolumn{3}{c}{Probe area}																\\[-0.5cm]                                                                   
				&	\multicolumn{1}{c}{10 kyr}		&	\multicolumn{1}{c}{100 kyr}	&	\multicolumn{1}{c}{1 Myr}		\\
	\hline\\[-1.5cm]
C$^+$		&	\mediummark	&	\mediummark		&	\cmark		\\[-0.6cm]
N$_2$H$^+$	&	\cmark		&	\peakmark			&	\barriermark	\\[-0.6cm]
HCO$^+$		&	\cmark		&	\cmark			&	\xmark		\\[-0.6cm]
HOC$^+$		&	\cmark		&	\cmark			&	\peakmark		\\[-0.6cm]
C			&	\bottommark	&	\barriermark		&	\cmark		\\[-0.6cm]
SO			&	\peakmark		&	\mediummark		&	\peakmark		\\[-0.6cm]
HNCO		&	\mediummark	&	\xmark			&	\peakmark		\\[-0.6cm]
C$_3$H$_4$	&	\cmark		&	\mediummark		&	\xmark		\\[0.0cm]
	\hline

  \end{tabular}
  \caption{Listing of the regions where selected individual species show the most promise as probes of \CR. The color-coding specifies how restricted the functional range of the probe is: none (green), small (yellow), moderate (orange), and severe (red). The symbols specify the nature of the abundance variation: ticks ($\checkmark$) indicate only one monotonic trend, tildes ($\sim$) signify some level of restriction, and bars ($\mid$) denote a symmetric arc-shaped distribution centered near $\sim 10^{-16}$ s$^{-1}$. 
  \label{tab:ind_regions}}
\end{table}

\section{Species pairs}\label{sec:pairs}
\begin{table*}[!htb]\scriptsize
 \centering
\begin{tabular}{lclclc}
\\	\hline	\hline\\[-0.7cm]
Species pair	&	Scores 	& Species pair	&	Scores		& Species pair	&	Scores					\\
\hline\\[-0.35cm]

\raisebox{\height-3px}{N$_2$H$^+$/C$_4$H}		&\greenscore{0.97}   \greenscore{0.82}   \greenscore{0.98}&  
\raisebox{\height-3px}{H$_3$O$^+$/C}			&\greenscore{0.70}   \yellowscore{0.63}  \greenscore{0.99}&  
\raisebox{\height-3px}{HC$_3$N/CH$_3$OH}		&\greenscore{0.82}   \greenscore{0.67}   \yellowscore{0.63}  \\

\raisebox{\height-3px}{N$_2$H$^+$/C$_2$H}		&\greenscore{0.97}   \greenscore{0.70}   \greenscore{0.98}&  
\raisebox{\height-3px}{C$_2$H/HNCO}			&\greenscore{0.98}   \orangescore{0.33}  \greenscore{0.99}&  
\raisebox{\height-3px}{C/NH$_2$D}				&\greenscore{0.77}   \yellowscore{0.63}  \greenscore{0.71}   \\

\raisebox{\height-3px}{N$_2$H$^+$/C}		    	&\greenscore{0.98}   \yellowscore{0.63}  \greenscore{0.99}&  
\raisebox{\height-3px}{N$_2$H$^+$/C$_3$H$_2$}	&\greenscore{0.98}   \greenscore{0.82}   \orangescore{0.50}& 
\raisebox{\height-3px}{SO/C$_4$H}				&\orangescore{0.33}  \greenscore{0.78}   \greenscore{0.99}   \\

\raisebox{\height-3px}{HCO$^+$/C}				&\greenscore{0.97}   \yellowscore{0.63}  \greenscore{0.98}&  
\raisebox{\height-3px}{C/CH$_3$OH}			&\greenscore{0.71}   \yellowscore{0.59}  \greenscore{0.99}&  
\raisebox{\height-3px}{C$_4$H/C$_3$H$_4$}		&\greenscore{0.78}   \greenscore{0.74}   \yellowscore{0.55}  \\

\raisebox{\height-3px}{C/C$_3$H$_4$}			&\greenscore{0.98}   \yellowscore{0.59}  \greenscore{0.90}&  
\raisebox{\height-3px}{H$_3$O$^+$/HNCO}		&\greenscore{0.98}   \yellowscore{0.59}  \greenscore{0.71}&  
\raisebox{\height-3px}{HNCO/C$_4$H}			&\greenscore{0.98}   \redscore{0.29}     \greenscore{0.79}   \\

\raisebox{\height-3px}{HOC$^+$/O}				&\greenscore{0.97}   \greenscore{0.97}   \orangescore{0.52}& 
\raisebox{\height-3px}{N$_2$H$^+$/HC$_3$N}		&\greenscore{0.98}   \greenscore{0.71}   \yellowscore{0.59}& 
\raisebox{\height-3px}{C/CN}					&\greenscore{0.73}   \greenscore{0.70}   \yellowscore{0.62}  \\

\raisebox{\height-3px}{CN/HNCO}				&\greenscore{0.98}   \orangescore{0.41}  \greenscore{0.99}&  
\raisebox{\height-3px}{N$_2$H$^+$/O}			&\greenscore{0.98}   \greenscore{0.74}   \yellowscore{0.55}& 
\raisebox{\height-3px}{HNCO/CH$_3$OH}		&\greenscore{0.99}   \orangescore{0.39}  \greenscore{0.67}   \\

\raisebox{\height-3px}{HOC$^+$/HNCO}			&\greenscore{0.98}   \greenscore{0.67}   \greenscore{0.71}&  
\raisebox{\height-3px}{N$_2$H$^+$/CS}			&\greenscore{0.98}   \greenscore{0.67}   \yellowscore{0.62}& 
\raisebox{\height-3px}{HOC$^+$/CS}			&\greenscore{0.97}   \greenscore{0.86}   \redscore{0.20}     \\

\raisebox{\height-3px}{N$_2$H$^+$/HNCO}		&\greenscore{0.99}   \greenscore{0.71}   \greenscore{0.67}&  
\raisebox{\height-3px}{SO/HC$_3$N}			&\greenscore{0.97}   \greenscore{0.67}   \yellowscore{0.63}& 
\raisebox{\height-3px}{C$^+$/N$_2$H$^+$}		&\greenscore{0.78}   \yellowscore{0.58}  \greenscore{0.67}   \\

\raisebox{\height-3px}{HCO$^+$/HNCO}			&\greenscore{0.98}   \yellowscore{0.59}  \greenscore{0.79}&  
\raisebox{\height-3px}{HCO$^+$/HC$_3$N}		&\greenscore{0.97}   \greenscore{0.70}   \yellowscore{0.59}& 
\raisebox{\height-3px}{H$_3$O$^+$/HC$_3$N}		&\greenscore{0.73}   \greenscore{0.70}   \yellowscore{0.59}  \\

\raisebox{\height-3px}{C/CO}					&\greenscore{0.77}   \yellowscore{0.59}  \greenscore{0.98}&  
\raisebox{\height-3px}{DCO$^+$/N$_2$H$^+$}		&\greenscore{0.97}   \greenscore{0.97}   \orangescore{0.32}& 
\raisebox{\height-3px}{C$^+$/HNCO}			&\greenscore{0.98}   \orangescore{0.31}  \greenscore{0.71}   \\

\raisebox{\height-3px}{HOC$^+$/C}				&\greenscore{0.97}   \yellowscore{0.63}  \greenscore{0.74}&  
\raisebox{\height-3px}{NH$_2$D/HC$_3$N}		&\yellowscore{0.66}  \greenscore{0.70}   \greenscore{0.82}&  
\raisebox{\height-3px}{C$^+$/O}				&\redscore{0.28}     \greenscore{0.74}   \greenscore{0.98}   \\

\raisebox{\height-3px}{N$_2$H$^+$/CN}			&\greenscore{0.97}   \yellowscore{0.58}  \greenscore{0.78}&  
\raisebox{\height-3px}{HOC$^+$/HC$_3$N}		&\greenscore{0.89}   \greenscore{0.70}   \yellowscore{0.55}& 
\raisebox{\height-3px}{HNCO/C$_3$H$_2$}		&\greenscore{0.90}   \orangescore{0.35}  \greenscore{0.71}   \\

\raisebox{\height-3px}{N$_2$H$^+$/SiO}			&\greenscore{0.97}   \greenscore{0.77}   \yellowscore{0.59}& 
\raisebox{\height-3px}{CO/HNCO}				&\greenscore{0.98}   \orangescore{0.43}  \greenscore{0.71}&  
\raisebox{\height-3px}{HCO$^+$/O}				&\greenscore{0.78}   \greenscore{0.70}   \redscore{0.27}     \\

\raisebox{\height-3px}{C$_2$H/HNCO}			&\greenscore{0.98}   \orangescore{0.33}  \greenscore{0.99}&  
\raisebox{\height-3px}{C/NH$_2$D}				&\greenscore{0.77}   \yellowscore{0.63}  \greenscore{0.71}&               &   \\
\hline
\end{tabular}
\caption{Listing of species pairs in this study that show, in at least two out of the three time steps, a high probing score corresponding to the top quartile of the total distribution. The species pairs are sorted by their mean probing score. The green scores represent the top quartiles with scores $\geq$ 0.66, the yellow scores represent the second quartiles with scores $\geq$ 0.53, the orange scores represent the third quartile with scores $\geq$ 0.31, and the red scores represent any non-zero values. \label{tab:ProbingScore_MolPairs}}
\end{table*}

If we look at pairs of species as probes, it is possible that they can remove restrictions or troublesome behaviors that individual species might have. Because forming pairs based on the selected species in Table~\ref{tab:selected_mols} will create a large number of pairs, we will focus our discussion here on the species pairs with probe scores from the top quartile (scores $\geq 0.66$) in at least two of the time steps. These are listed in Table~\ref{tab:ProbingScore_MolPairs}, sorted by their mean score. From the probe scores we can immediately see that the species pairs perform much better than than the individual species. Many of these pairs have their lowest score at 100 kyr, but there are also many examples of species that have their lowest probe score at 1 Myr. Only a few exceptions have problems at 10 kyr: NH$_2$D/HC$_3$N, SO/C$_4$H and C$^+$/O, but they perform much better at the other time steps. 

From the more than 200 species pairs generated, only 44 have at least 2 green probe scores amongst the 3 time steps. Not surprisingly, the majority of these pairs include an ion (22/44). The most commonly listed ion is N$_2$H$^+$ (12/44), which follows from it being one of the most promising individual probes. After this, HCO$^+$ and HOC$^+$ (4/44 and 5/44, respectively) are significantly less common. Amongst the neutrals, HNCO (12/44) is most common, mostly owing to its large abundance variation. It is closely followed by C (8/44) and HC$_3$N (6/44). With the exception of HC$_3$N, these belongs to the best performing individual species that we identified in Section~\ref{sec:ind_tracers}. 

As an individual tracer, the mean score of HC$_3$N is pulled down because it is not affected by the increasing \CR~at 10 kyr. Otherwise HC$_3$N performs well with green scores at other time steps. When paired with other species that are strongly affected by \CR~at 10 kyr, we get a much more promising probe. 

Looking at the size of the species pairs, it appears that small differences of 1 $-$ 2 atoms result in the best probes. Looking at differences in atomic weight between the two species, there is a grouping around 13$-$17 in atomic mass, identified as the difference between one heavier element (C, N, O), and possibly also a lighter element (H, D). These small differences suggest that the formation process for either partner of a species pair should have a length difference of one, or at most two, steps. However, the two individual species in the pairs often are formed from different formation pathways. The pairs N$_2$H$^+$/C$_4$H and N$_2$H$^+$/C$_2$H, are excellent examples of this, where one species belongs to nitrogen chemistry, and the other to carbon chemistry. In short: similarity is preferential, but they should not be too similar.

\begin{figure*}[!tb]
\centering
\includegraphics[width=1.00\textwidth]{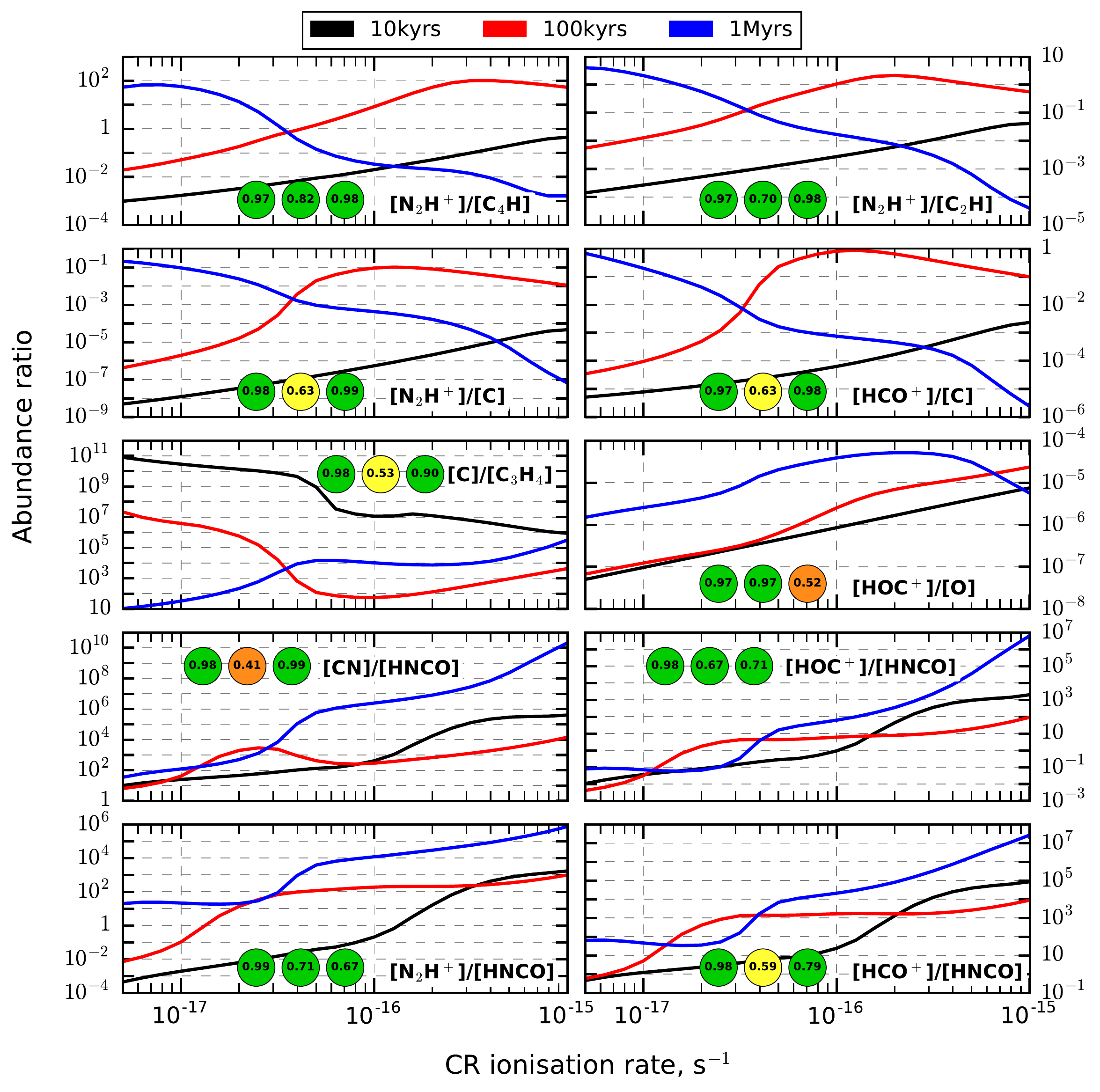}
\caption{Abundance as a function of $\zeta_{CR}$ for differently behaving molecules at times 0.01 (black line), 0.1 (red line) and 1.0 Myr (blue line). Green scores belong to the top quartiles with a score $\geq 0.66$, yellow for the second quartiles with scores $> 0.53$, orange for third quartile with scores $> 0.31$, and red for any non-zero values.
}
\label{fig:showcasePairs}
\end{figure*} 

In Figure~\ref{fig:showcasePairs} we plot the abundances of the top 10 scoring species pairs from Table~\ref{tab:ProbingScore_MolPairs}. All of these have a clear monotonic trend at 10 kyr, and the difference between these is their behavior at later times. Most of them have their lowest probe score at 100 kyr, an issue that it shares also with the individual probes. Only four pairs have green scores in all three time steps: N$_2$H$^+$/C$_4$H, N$_2$H$^+$/C$_2$H, HOC$^+$/HNCO, and N$_2$H$^+$/HNCO. 

The two topmost pairs N$_2$H$^+$/C$_4$H and N$_2$H$^+$/C$_2$H show great promise as probes at all three time steps. Their greatest limitation is still at 100 kyr at \CR~$\gtrsim 10^{-16}$ s$^{-1}$, meaning they are still great probes for more than half the \CR~range. The other two fully green pairs HOC$^+$/HNCO and N$_2$H$^+$/HNCO both have steady monotonic increases at 10 kyr, but at later times there are some plateaus that affect the functional range, and limit the probe scores. This does not change the fact that also these are very capable \CR~probes. 

One particular species pair does not have its lowest probe score at 100 kyr: HOC$^+$/O. It is limited to \CR~$\lesssim 2\times 10^{-16}$ s$^{-1}$ and along with a limited functional range, it receives a low score at 1 Myr. N$_2$H$^+$/HNCO also has its lowest scores at 1 Myr. But it produces high probe scores for all time steps, only limited in functional range by plateaus. 

The remaining pairs are limited by arc-shaped distributions or extensive plateaus at 100 kyr, and either have a monotonic trend, or small limitations by plateaus or an arc-shaped distribution skewed toward the high end of the \CR~range. Even with these limitations, they are capable as probes, as long as one knows enough about the targets to avoid the problematic time scales and areas in the \CR~range.

We can clearly see that there are promising probes amongst these species pairs, and in Table~\ref{tab:pairs_regions} we list a summary of how well the most promising species pairs perform. Given that our species selection in Table~\ref{tab:selected_mols} is largely driven by observations, most of the abundance ratios listed in Table 4 can be observed using (sub)millimeter--wavelength telescopes. Species like C$_4$H, C$_3$H$_2$, and HOC$^+$ might require significant integration times, but they are generally observable. The situation is different for ratios involving atoms, though. Observations of C at 492~GHz from the ground are only possible in exceptionally high and dry sites. 

High-spectral resolution observations of the ground-state fine structure of neutral atomic oxygen at 63 $\mu$m wavelength (4.7 THz) are now routinely possible with the German Receiver for Astronomy at Terahertz Frequencies \citep[GREAT;]{Risacher2016} on board the Stratospheric Observatory for Infrared Astronomy (SOFIA). The recent SOFIA observations cover this line in the dense \citep{Leurini2015} and diffuse ISM \citep{Wiesemeyer2016}. 

\begin{table}[!tb]\small
\centering
  \begin{tabular}{cccc}
	\hline	\hline\\[-1.2cm]
Species			&	\multicolumn{3}{c}{Probe area}																\\[-0.4cm]                                                                   
				&	\multicolumn{1}{c}{10 kyr}		&	\multicolumn{1}{c}{100 kyr}	&	\multicolumn{1}{c}{1 Myr}		\\
\hline\\[-1.2cm]
N$_2$H$^+$/C$_4$H	&	\cmark		&	\peakmark		&	\peakmark		\\[-0.45cm]
N$_2$H$^+$/C$_2$H	&	\cmark		&	\peakmark		&	\cmark		\\[-0.45cm]
N$_2$H$^+$/C			&	\cmark		&	\mediummark	&	\cmark		\\[-0.45cm]
HCO$^+$/C			&	\cmark		&	\mediummark	&	\cmark		\\[-0.45cm]
C/CH$_3$OH			&	\xmark		&	\barriermark	&	\cmark		\\[-0.45cm]
HNCO/C$_4$H			&	\peakmark		&	\xmark		&	\peakmark		\\[-0.45cm]
C/CO				&	\cmark		&	\barriermark	&	\cmark		\\[-0.45cm]
HOC$^+$/	O			&	\cmark		&	\cmark		&	\barriermark	\\[-0.45cm]
N$_2$H$^+$/C$_3$H$_2$&	\cmark		&	\cmark		&	\bottommark	\\[-0.45cm]
HOC$^+$/HNCO		&	\cmark		&	\bottommark	&	\mediummark	\\[0.05cm]
	\hline
  \end{tabular}
  \caption{Listing of the regions of the most promising species pairs based on their average probe scores. The color-coding specifies how restricted the functional range of the probe is: no (green), small (yellow), moderate (orange), and severe (red). The symbols specify the nature of the abundance variation: ticks ($\checkmark$) indicate only one monotonic trend, tildes ($\sim$) signify some level of restriction, and bars ($\mid$) represent an arc-shaped distribution.
\label{tab:pairs_regions}}
\end{table}

\section{Conclusions}\label{sec:conclusions}
We have looked at the potential of individual and pairs of species as probes for \CR~in dense clouds. The study is limited to species commonly observed in large-scale surveys and for \CR~=~$5\times 10^{-18} - 1\times10^{-15}$~s$^{-1}$ at the time steps 0.01, 0.1, and 1.0 Myr. Our study is a pilot intended to be expanded to include a larger sample of species to prepare for the new generation of telescopes and future surveys. 

Ions are strongly linked to an increase in \CR, but molecular ions suffer from molecular dissociation at high \CR~$\gtrsim 10^{-16}$ s$^{-1}$ and 1~Myr. This means that primarily atoms are increasing in abundance beyond this point, and as we look at later time steps, the molecular dissociation is dominating at even lower \CR. Atomic ions continuously increase with \CR, and even more so when the molecular dissociation is dominating. Because the abundance of some atomic ions, in particular H$^+$, C$^+$, He$^+$, as well as those of some abundant molecular ions (e.g. H$_3^+$, HCO$^+$, H$_3$O$^+$), are very reactive, the atomic/molecular as well as ion/neutral ratios will affect the chemistry, and the former in particular could possibly also be used as a probe. 

The probe candidates show a range of behaviors depending on their chemistry and in many cases it is the complexity of the species that determines how sensitive they will be to \CR. It controls the abundances of atomic and molecular ions as well as electrons, which are the driving components of the destructive pathways. Large, complex molecules with long formation pathways will decrease or even dilute the effects of the steadily increasing \CR, while abundances are low because reforming them takes a long time. 

In order to infer how great an individual or pair of species can probe \CR, we identify different trends in the abundance variations and use them to calculate a probe score. It is based on the multiplication of two factors: one relating to the $functional$ range of the probe, as in how much of the \CR~range it can be used as a probe, and one factor relating to the magnitude of abundance $variation$ that the species experience over the studied \CR~range. We find that all probe candidates investigated in this study have limitations. That is, none of the individual or pairs of species are perfect probes for the complete investigated \CR~range and at all three time steps, but a few come very close with only small limitations. 

Amongst individual species, the ions N$_2$H$^+$, HCO$^+$ and HOC$^+$ are the most capable probes. HNCO is a particularly interesting molecule, as it shows an incredibly large range of abundance variation at 1 Myr of almost eight orders of magnitude. This sensitivity is caused by its formation being slow and largely associated with surface chemistry. This may make it a useful probe for more distant targets, as it will not be as sensitive to large errors that can be expected for distant objects. 

Species pairs show more promise. Most of them show a monotonic trend with an increasing \CR~at 10 kyr and 1 Myr, while many suffer from restrictions during the transitional period at 100 kyr, where many trends are reversed as molecular dissociation begins to affect the chemistry. 

Amongst the pairs, there are a few that stand out. N$_2$H$^+$/C$_4$H only shows restrictions at 100 kyr for high \CR~$\gtrsim 3\times10^{-16}$ s$^{-1}$ and at 1 Myr for low \CR~$\lesssim 8\times 10^{-18}$ s$^{-1}$. N$_2$H$^+$/C$_2$H on the other hand has no restrictions at 1 Myr, but is more limited at 100 kyr to \CR~$\lesssim 10^{-16}$ s$^{-1}$. N$_2$H$^+$/C$_2$H, however, on the other hand has no restriction at 1 Myr, but a larger restriction at 100 kyr due to the peak of the arc being shifted to lower \CR. 

Many species pairs are restricted or not very useful as probes at 100 kyr due to low variations or the trends not being monotonic. But there is an equal number of species pairs that have their lowest probe score at 1 Myr. However, the highest scoring species pairs are almost exclusively belong to the former type. 

A number of pairs with HNCO as a partner have the very particular property that their abundance varies by almost eight orders of magnitude at 1 Myr. This can be particularly useful in studies of distant targets, where larger abundance variations could improve the accuracy of the estimated \CR. 

We have summarized the functional ranges of individual species and species pairs, so that future surveys and observations can target the right probes for the right environments. While this study focused on dense clouds, at least some of these probes are likely also applicable in diffuse clouds and other astrochemical environments. This can be investigated in a future study, and the current one could also be extended to include more species to prepare for the next generation of observational facilities.

\acknowledgments
This research made use of NASA's Astrophysics Data System. 

\software{ALCHEMIC \citep{2010A&A...522A..42S}}

\bibliography{main}{}

\end{document}